\documentclass[runningheads,a4paper]{llncs}

\usepackage{amssymb}
\usepackage{graphicx}
\usepackage{url}

\urldef{\mailsa}\path|adrian.paschke@inf.fu.berlin.de|
\newcommand{\keywords}[1]{\par\addvspace\baselineskip
\noindent\keywordname\enspace\ignorespaces#1}

\begin{document}

\mainmatter  % start of an individual contribution

\title{OntoMaven - Maven-based Ontology Development and Management of Distributed Ontology Repositories}

\titlerunning{OntoMaven - Maven-based Ontology Development}

\author{Adrian Paschke}
\authorrunning{OntoMaven - Maven-based Ontology Development}

\institute{Corporate Semantic Web, Institute of Computer Science,\\
Koenigin-Luise-Str. 24, 14195 Berlin, Germany\\
\mailsa\\
\url{http://www.corporate-semantic-web.de}}

\toctitle{OntoMaven: }
\tocauthor{Maven-based Ontology Development and Management of Distributed Ontology Repositories}
\maketitle

\begin{abstract}
In collaborative agile ontology development projects support for modular reuse of ontologies from large existing remote repositories, ontology project life cycle management, and transitive dependency management are important needs. The Apache Maven approach has proven its success in distributed collaborative Software Engineering by its widespread adoption. The contribution of this paper is a new design artifact called OntoMaven. OntoMaven adopts the Maven-based development methodology and adapts its concepts to knowledge engineering for Maven-based ontology development and management of ontology artifacts in distributed ontology repositories.

\end{abstract}

\section{Introduction}
Sharing and reusing knowledge in ontology-based applications is one of the main aims in the Semantic Web as well as the Pragmatic Web\footnote{\scriptsize{\url{http://www.pragmaticweb.info}}} \cite{DBLP:conf/ruleml/WeigandP12,DBLP:journals/ijait/PaschkeB11,DBLP:conf/icpw/PaschkeBKC07}, which requires the support of distributed ontology management, documentation, validation and testing.   As pointed out in \cite{DeNicola:2009:SEA:1467085.1467122,PaCoHe10} such ontology development life cycles have a similar structural and logic complexity as distributed software development projects. Agile ontology life cycle management methodologies, such as COLM\footnote{\scriptsize{\url{http://www.corporate-semantic-web.de/colm.html}}} \cite{RoeHe09}, require the collaboration of knowledge engineers and domain experts. Ontologies are developed and maintained in an iterative and distributed way, which requires the support of versioning \cite{conf/gi/Luczak-RoschCPRT10,PA155} and modularization \cite{CLHP2009a,Coskun2012}. Moreover, new aspect-oriented ontology development approaches \cite{SchPa13} enable weaving of  cross-cutting knowledge concerns into the main ontology model, which requires meta-level descriptions of ontology aspects and management of distributed knowledge models.

In this work we adapt a highly successful method and tool in distributed software engineering project management, namely Apache Maven\footnote{\scriptsize{\url{http://maven.apache.org/}}}, for the Maven-based management of distributed ontology repositories. Maven is tool for the (distributed) project management and quality assurance in software engineering projects. The goal of Maven is to automate recurring tasks in software development projects such as management of software artifact in distributed remote and local repositories, versioning, dependency management, documentation, testing, packaging and deployment.

We follow a Design Science research methodology and develop a novel design artifact, called OntoMaven\footnote{\scriptsize{\url{http://www.corporate-semantic-web.de/ontomaven.html}}}, as a relevant and original contribution in distributed ontology engineering. The approach is grounded in the rigorous methods of Maven's engineering approach, which has already proven its value and success in software engineering. The OntoMaven approach supports ontology engineering in the following ways:

\begin{itemize}
\item OntoMaven remote repositories enable distributed publication of ontologies as \textbf{ontology development artifacts} on the Web, including their metadata information about life cycle management, versioning, authorship, provenance, licensing, knowledge aspects, dependencies, etc.
\item OntoMaven local repositories enable the reuse of existing ontology artifacts in the users' local ontology development projects.
\item OntoMaven's support for the different development phases from the design, development to testing, deployment and maintenance provides a flexible life cycle management enabling iterative agile ontology development methods, such as COLM \cite{RoeHe09}, with support for collaborative development by, e.g., OntoMaven's dependency management, version management, documentation and testing functionalities, etc.
\item OntoMave plug-ins provide a flexible and light-weight way to extended the OntoMaven tool with existing functionalities and tools, such as semantic version management (e.g., SVont - Subersion for Ontologies \cite{conf/gi/Luczak-RoschCPRT10,PA155}), semantic documentation (e.g., SpecGen Concept Grouping \cite{Coskun2012}), dependency management of aspect-oriented ontology artifacts (e.g. \cite{SchPa13}), automated testing (e.g., with the W3C OWL test cases and external reasoners such as Pellet), etc.
\item Maven's API allows easy integration of OntoMaven into other ontology engineering tools and their integrated development environments (IDE).
\end{itemize}

The further paper is structured as follows: Section \ref{relatedwork} describes related work. Section \ref{OntoMavenConcept} describes the design of OntoMaven based on Maven's remote and local repositories, the Project Object Model (POM), plug-ins, and important functional concepts of the solution approach for OntoMaven ontology development. Section \ref{proof-of-concept} proves the feasibility of the proposed concepts with a proof-of-concept implementation of the OntoMaven design artifact. Section \ref{evaluation} compares the OntoMaven functionalities to the tool support of the major existing ontology engineering tools, namely Protege, Swoop and Top Braid Composer. This highlights the strengths of OntoMaven with the open approach to model, manage, and reuse ontology (modules) as \textbf{ontology development artifacts} including their metadata descriptions in the POM, dependency management, versioning, documentation, etc. Finally, section \ref{conclusion} summarizes the current OntoMaven work and discusses future research.

\section{Related Work}
\label{relatedwork}

There are many existing ontology engineering methodologies and ontology editors available. With its Maven-based approach for structuring the development phases into different goals providing different functionalities during the development project's life cycle, OntoMaven supports in particular agile ontology development methods, such as RapidOWL \cite{DBLP:conf/wetice/Auer06} and COLM \cite{RoeHe09}, as well as development methods which are inherently based on modularization such as aspect-oriented ontology development \cite{SchPa13}.

According to \cite{KhiMue10} the most popular ontology editors supporting the Semantic Web ontology languages (RDF, RDFS, OWL, SWRL) are Protege\footnote{\scriptsize{\url{http://protege.stanford.edu/}}}, Swoop\footnote{\scriptsize{\url{http://www.mindswap.org/2004/SWOOP/}}}, and Top Braid Composer\footnote{\scriptsize{\url{http://www.topquadrant.com/products/TB_Composer.html}}}. Other editors support, e.g., visual ontology modeling such as Thematix Visual Ontology Modeler (VOM)\footnote{\scriptsize{\url{http://thematix.com/tools/vom/}}}, which enables UML-based ontology modeling based on the OMG Ontology Definition Metamodel (OMG ODM\footnote{\scriptsize{\url{http://www.omg.org/spec/ODM/}}}), or lightweight domain specific vocabulary development, such as Leone\footnote{\scriptsize{leone - \url{http://www.corporate-semantic-web.de/leone.html}}} \cite{PA155}. While the focus of OntoMaven is on supporting the backend-functionalities in ontology development projects, the focus of these editors is on the support of ontology modeling / representation with a user interface. Non of them is based directly on the Apache Maven concepts and its methods. While all of the editors also provide support for ontology repositories and reuse of existing ontologies by imports, OntoMaven mainly differs in the approach how it manages and declaratively describes the development phases, goals, and artifacts in a Maven Project Object Model (POM). Further implementation-specific and plug-in-specific differences are in the underlying details of the provided functionalities of OntoMaven such as POM-based dependency management, semantic versioning, semantic documentation etc. For a comparison see the evaluation section \ref{evaluation}.

The W3C Wiki lists several existing ontology repositories\footnote{\scriptsize{\url{http://www.w3.org/wiki/Ontology_repositories}}}. Further ontology repositories are, e.g., COLORE\footnote{\scriptsize{\url{http://stl.mie.utoronto.ca/colore/}}} for ontologies written in the ISO Common Logic (CL) ontology languages and Ontohub\footnote{\scriptsize{\url{http://ontohub.org/}}} which maintains a set of heterogenous ontologies. The current focus of these projects is on collection and listing of existing ontologies. Apart from simple search functionalities there is no support for repository-based ontology development which is the focus of OntoMaven and OntoMaven repositories.

New standardization efforts such as OMG Application Programming Interfaces for Knowledge Bases (OMG API4KB)\footnote{\scriptsize{\url{www.omgwiki.org/API4KB/}}} and OMG OntoIOP aim at the accessability and interoperability of heterogenous ontologies via standardized interfaces and semantic transformations defined on the meta-level of the ontology models, e.g., by the Distributed Ontology Language (DOL) \cite{DBLP:journals/corr/abs-1204-5093}. These approaches do not address ontology engineering directly, but can provide a standardized repository back-end for OntoMaven ontology development projects.

\section{OntoMaven's Design and Concept }
\label{OntoMavenConcept}

This section describes the approach and the concepts of the new design artifact, called OntoMaven, which adapts Apache Maven for the Maven-based ontology development and management of distributed OntoMaven ontology repositories. Maven is not just an automated build tool but also supports software artifact management and quality assurance in software projects. The main required functionalities provided by Maven are:

\begin{itemize}
\item source code compilation
\item dependency management
\item testing with test suites
\item automated documentation and reporting
\item installation and deployment of generated code
\end{itemize}

The main design concepts of Maven are:

\begin{itemize}
\item The Project Object Model (POM) is the main declarative XML description for managing a project and its development artifacts. Based on the instructions in a POM file Maven automates the different project goals in the life cycle of a software development project.
\item Maven plug-ins implement the functionality of the different Maven goals and lead to a modular and extensible architecture of Maven. The plug-ins are executed by Maven using the descriptions in the POM file. Maven has three predefined life cycles, namely the \emph{Clean} life cycle, which cleans the project, the \emph{Default} life cycle, which processes, builds, tests and installs locally or deploys remotely, and the \emph{Site} life cycle, which reports, documents and deploys the created HTML documentation, e.g. on a Web server.
\item Maven local and remote repositories manage the used plug-ins and artifacts including support for versioning and dependency management. The general approach is that libraries of existing software frameworks and Maven plug-ins which are required in a software development project are downloaded from distributed remote repositories to the local Maven repository so that Maven can work with these artifacts locally during the development. This distributed approach supports sharing and reuse of existing software artifacts. The information about the used artifacts and their remote addresses (typically a URL) as well as dependency information are described in the POM file of a project. The downloaded artifacts have their own POM files in order to support e.g. transitive dependencies.
\end{itemize}

In the following subsections we adapt the main concepts of Maven, so that they can be used in ontology development and ontology life cycle management. In particular, we focus on the (distributed) management of knowledge artifacts (ontologies / ontology modules) and their versioning, import and dependency management, documentation, and testing.

\subsection{Management and Versioning of Ontology Artifacts}
\label{versioning}

One of the design patterns in ontology engineering is the reuse of existing ontologies and ontology modules. Finding ontologies on the Web is supported, e.g., by specialized ontology search engines such as Swoogle\footnote{\scriptsize{\url{http://swoogle.umbc.edu/}}} and Watson\footnote{\scriptsize{\url{http://kmi-web05.open.ac.uk/WatsonWUI/}}}. Since such found ontologies typically cannot be used directly, but need to modularized, refactored and enhanced, before they can be reused in an ontology development project, there is  need for versioning and life cycle management of such ontologies. Furthermore, combinations with other existing ontologies (by ontology matchmaking and alignment) might lead to transitive dependencies which need to be described and managed. OntoMaven therefore adopts Maven's artifact concept. It describes and manages ontologies as ontology artifacts in a Maven Project Object Model (POM). The typical steps to add an ontology (module) as an OntoMaven artifact to a POM are:

\begin{enumerate}
  \item Find ontology module(s)
  \item Select the right module and version
  \item Analyse and resolve dependencies of the modules
  \item Declaratively describe the ontology artifact in a POM
\end{enumerate}

Many ontology languages support imports or integration of distributed ontologies. For instance, the W3C Web Ontology Language (OWL) therefore has a specialized \texttt{owl:import} statement and the ISO Common Logic standard supports modularized imports by a segregation semantics which distinguishes the universe of discourse of the main ontology from the segregated universe of discourse of the imported ontology.

Typical recurring tasks which are automated by OntoMaven are in such modular import and reuse scenarios are,

\begin{itemize}
  \item check the existence of the imported ontology (module) referenced by the defined URI in the import statement (and find alternative URLs from pre-configured repositories if the ontology does is not found at the import URI).
  \item management of ontologies / ontology modules as ontology artifacts in Maven repositories including their metadata descriptions such as versioning information.
  \item download of ontology artifacts from remote repositories (including transitive imports) to a local development repository in order to support offline development of ontologies
\end{itemize}

Another important aspect in the agile and collaborative development of ontologies is the support for version management. Typical requirements are maintaining consistency and integrity, as well as provenance data management (e.g. authorship information) throughout the version history. Compared to version management solutions in software engineering which just maintain syntactic versions on the level of code line differences, the difficulty in knowledge engineering is that the versions and their differences need to be managed on a semantic level. Different syntactic versions of an ontology knowledge model might still have the same semantic interpretation. That is, a semantic version management system needs to compute the semantic difference, e.g., to detect and resolve version conflicts. The approach in OntoMaven is based on the ontology versioning tool \emph{SVont}\footnote{\scriptsize{\url{http://www.corporate-semantic-web.de/svont.html}}}, which is an extensions of the version management tool Subversion. \cite{conf/gi/Luczak-RoschCPRT10,PA155}

\subsection{Import and Dependency Management}
\label{dependencyManagement}

OntoMaven adopts the dependency management of Maven by describing the dependencies in an ontology development project from existing ontology artifacts in a POM. This is illustrated in the following listing from a POM example:

\begin{scriptsize}
\begin{verbatim}
<project>
  ...
  <dependencies>
    <dependency>
        <groupId>de.onto.maven</groupId>
        <artifactId>TimeOntologie</artifactId>
        <version>1.0</version>
    </dependency>
...
</dependencies>
</project
\end{verbatim}
\end{scriptsize}

The listing describes a dependency of an ontology artifact identified by the ID \texttt{TimeOntologie} version \texttt{1.0} which belongs to the group \texttt{de.onto.maven}.

It is possible to define multiple repositories in which OntoMaven will look for dependent ontology artifacts. If the defined ontology artifact is not found in the repository, users will be informed. They can start a search with the ontology name and URI defined in the artifact description in configured ontology search engines. OntoMaven supports Swoogle and Watson. The found ontologies will can be added as new ontology artifacts to an OntoMaven repository.

\subsection{Documentation}
Despite first automated ontology matching and alignment approaches developing high quality ontologies still remains a manual knowledge modeling effort, where domain experts work together with knowledge engineers. Documentation is an important step which facilitates this work and in particular makes maintenance and reuse easier. The typical distinction is into \emph{user documentation} and \emph{technical documentation}. While the former supports the users of an ontology, e.g. in their task to populate the ontology with instance data, the latter, technical documentation, supports the ontology developer.

There exist several ontology documentation tools such as OWLDoc\footnote{\scriptsize{\url{http://docpp.sourceforge.net/}}}, VocDoc\footnote{\scriptsize{\url{http://kantenwerk.org/vocdoc /}}}, or SpecGen\footnote{\scriptsize{\url{https://github.com/specgen/specgen}}}, which document ontologies on a technical level (much like tools such as JavaDoc in Java programming). Unfortunately, the lack of good documentation in the published ontologies on the Web makes reuse difficult, because the analysis and decision process on the applicability of a candidate ontology becomes very time-consuming. Therefore, additional automated support needs to be provided, e.g., for analysing larger ontologies on an abstract level by  creating concept groupings which reduce the complexity of the ontology model. This groupings and summarizations of concepts provides the reader with an easier way to understand the ontology vocabulary. Typically such concept groups are additionally presented in easy to understand visualization formats, e.g. by tools such as OWLViz\footnote{\scriptsize{\url{http://www.co-ode.org/downloads/owlviz /}}}, OntoGraf\footnote{\scriptsize{\url{http://protegewiki.stanford.edu/wiki/OntoGraf}}}, Sonivis OWL plugin\footnote{\scriptsize{\url{http://www.corporate-semantic-web.de/ontology-modularization-framework.html}}}, SOVA\footnote{\scriptsize{\url{http://protegewiki.stanford.edu/wiki/SOVA}}}, TGVizTab\footnote{\scriptsize{\url{http://users.ecs.soton.ac.uk/ha/TGVizTab /}}}, or as UML models, e.g. by VOM\footnote{\scriptsize{\url{http://thematix.com/tools/vom/}}}. Such concept grouping and visualisations can be used in the user documentation of an ontology. \cite{Coskun2012}

Maven supports the documentation phase and provides goals for creating and publishing automated reports. In OntoMave we make use of SpecGen and the SpecGen extension\footnote{\scriptsize{\url{http://www.corporate-semantic-web.de/concept-grouping.html}}} for automated concept grouping, in order to create the technical and user documentation in an OntoMaven plugin which is executed by the \texttt{mvn site} command.

\subsection{Testing}
Testing is an important phase in the ontology life cycle. In particular, in agile iterative development processes testing allows detecting inconsistencies, anomalies, improper design, as well as validation against, e.g., the intended results of domain experts' competency questions which are represented as ontology test cases. Maven supports a testing phase in which automated tests are executed and the results are reported by the Maven command \texttt{mvn test}.

The W3C OWL recommendation\footnote{\scriptsize{\url{http://www.w3.org/TR/owl-test/}}} defines a collection of test cases with different test types and tests. As standard test types OntoMaven by default supports the W3C OWL test cases \emph{syntax checker}, \emph{consistency checker}, and \emph{entailment test}. The produced test results are compliant to the W3C recommendation and the created test reports show if the ontology model is \texttt{consistent}, \texttt{inconsistent} or if the result is \texttt{unknown}. Further test types can be implemented as Maven plug-ins and added to the OntoMaven projects test suites by the user.

\section{Proof-of-Concept Implementation - OntoMaven PlugIns}
\label{proof-of-concept}

The implementation of OntoMaven extends and adapts Maven \cite{kilic13}, so that it supports the management of ontology modules in Maven repositories. This section describes how the OntoMaven approach and the above described concepts are implemented using \emph{Maven repositories} and the \emph{Maven plug-in} extension mechanism. A Maven plug-in is a collection of one or more goals. For instance, the plug-in \texttt{archetype} implements the goals \texttt{create} and \texttt{generate} which create a Maven project. The implementation of a Maven plug-in is done in an \emph{Maven Plane Old Java Object (MOJO)}. Maven supports the automated generation of a Mojo project with the goal \texttt{generate} in the Maven plugin \texttt{Archetype}:

\begin{scriptsize}
\begin{verbatim}
Mvn archetype:generate -DinteractiveMode=false
-DarchetypeArtifactId=maven-archetype-mojo -DgroupId=[] -DartifactId=[]
\end{verbatim}
\end{scriptsize}

By defining the \texttt{groupId} and the \texttt{artifactID} the command creates a Mojo project with a Mojo class which is used for the plug-in implementation.

In the OntoMaven approach, the phases and goals, which the plug-in implements, are defined by JavaDoc annotations in the source code of the Mojo class. For instance, the following annotations define that the implemented plug-in is used in the phase \texttt{test} and that is has a goal called \texttt{test-syntax}:

\begin{scriptsize}
\begin{verbatim}
@phase test //plug-in used in test phase
@goal test-syntax // goal with the name "test-syntax"
\end{verbatim}
\end{scriptsize}

Parameters are used to configure the plug-in execution. For instance, the following code snippet defines a required parameter \texttt{compliancemode} with the default value \texttt{strict}:

\begin{scriptsize}
\begin{verbatim}
*@parameter expression = “compliancemode“
*default-value="strict"
*@required
\end{verbatim}
\end{scriptsize}

Such plug-in parameters can be configured in a POM.xml file or directly when calling a goal, e.g. \texttt{mvn ... -Dcompliancemode=strict}.
An implemented plug-in can be installed using Maven \texttt{mvn install} and the plug-in goals can be integrated into the POM.xml of an OntoMaven project, as the following example listing shows for the plug-in SVontPlugin and the goal semantic-diff:

\begin{scriptsize}
\begin{verbatim}
<build>
    <plugins>
        <plugin>
            <groupId>de.csw.ontomaven</groupId>
            <artifactId>SVontPlugin</artifactId>
            <version>1.0-SNAPSHOT </version>
            <executions>
              <execution>
               <goals>
                <goal>semantic-diff</goal>
               </goals>
              </execution>
            </executions>
        </plugin>
    </plugins>
</build>
\end{verbatim}
\end{scriptsize}

The following subsections provide further details about the proof-of-concept implementations of the main plug-ins in OntoMaven. We first describe the OntoMaven repositories which are the persistence and back-end layer for storing and managing ontologies.

\subsection{OntoMaven Repositories}
OntoMaven can use all Maven compliant repositories. One of the strengths of Maven is that is uses a folder structure following a standard folder layout for its repositories; sources are in \texttt{\${basedir}/src/main/java}, resources in \texttt{\${basedir}/src/main/resource}, tests in \texttt{\${basedir}/src/test}, classes in\\ \texttt{\${basedir}/target/classes}, and packaged libraries in \texttt{\${basedir}/target/}.

\begin{figure}
  \centering
  \includegraphics[width=8cm]{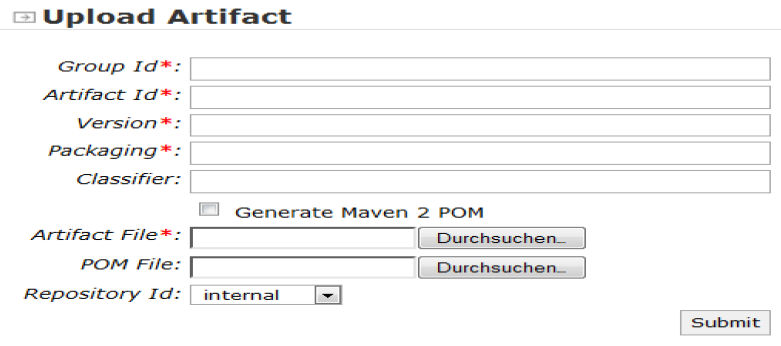}\\
  \caption{Archiva User Interface - Ontology Artifact Upload}\label{fig:ArchivaUpload}
\end{figure}

For the OntoMaven proof-of-concept implementation we adapted the Apache Archiva Build Artifact Repository Manager\footnote{\scriptsize{\url{http://archiva.apache.org/}}} as a managing tool providing a user interface for the OntoMaven repositories. It supports finding and managing OntoMaven artifacts. Figure \ref{fig:ArchivaUpload} shows the upload user interface.

Via this form an ontology can be uploaded to an OntoMaven repository together with its POM file. The artifact's metadata contains information about the group id, artifact id, version, packaging and optional additional classifier information. The POM provides all necessary information about the artifact and its dependencies. In OntoMaven these dependencies are used to describe (transitive) imports from an ontology, which are resolved by the OntoMvnImport plug-in (see subsection \ref{OntoMvnImportPlugIn}) and are defined in the ontology's POM file.

\begin{figure}
  \centering
  \includegraphics[width=8cm]{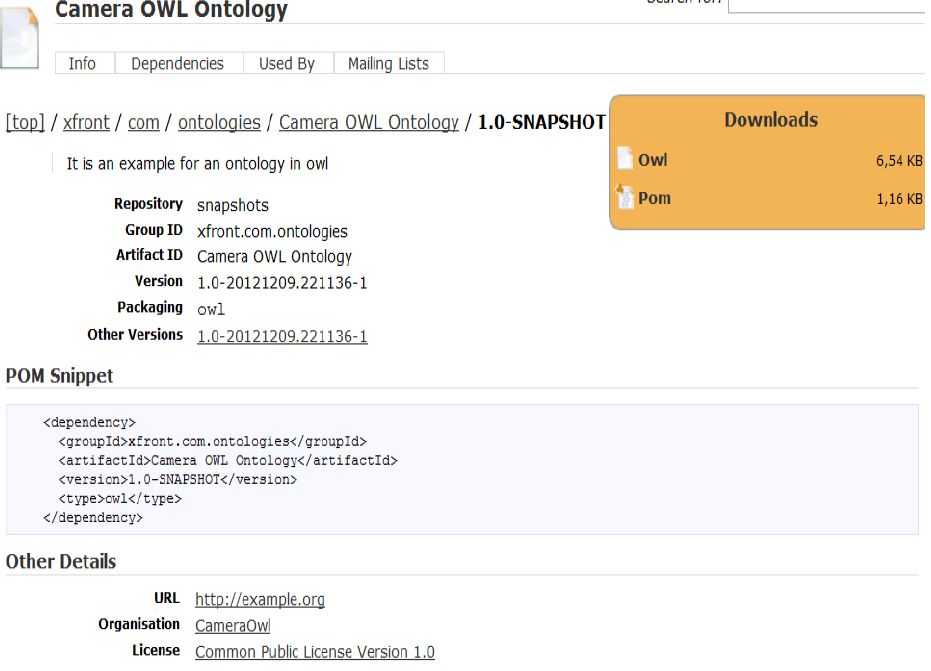}\\
  \caption{Archiva User Interface - Management of Ontology Artifact}\label{fig:ArchivaArtifactManagement}
\end{figure}

Figure \ref{fig:ArchivaArtifactManagement} gives an example of the Archiva user interface showing the management information of an ontology artifact called \texttt{Camera OWL Ontology}. Under the interface menu link \texttt{Dependencies} the dependencies of this ontology can be found.

Once managed in an online OntoMaven repository, an ontology artifact can be used in any OntoMaven ontology development project. The following listing gives an example how a remote repository can be configured and a dependency to an ontology artifact (here the Camera-OWL-Ontology) can be defined in the POM.xml document of a project.

\begin{scriptsize}
\begin{verbatim}
<profiles>
    <profile>
        <id>2</id>
        <activation>
            <activeByDefault>true</activeByDefault>
        </activation>
        <repositories>
            <repository>
                <snapshots>
                    <enabled>true</enabled>
                </snapshots>
                <id>snapshots</id>
                <name>OntoMaven Snapshot Repository</name>
                <url>www.corporate-semantic-web.de/repository/snapshots/</url>
            </repository>
        </repositories>
    </profile>
</profiles>

<dependencies>
    <dependency>
        <groupId>xfront.com.owl.ontologies</groupId>
        <artifactId>Camera-OWL-Ontology</artifactId>
        <version>1.0-SNAPSHOT</version>
        <type>owl</type>
    </dependency>
</dependencies>
\end{verbatim}
\end{scriptsize}

\subsection{OntoMvnImport}
\label{OntoMvnImportPlugIn}

This plug-in implements the imports of Ontologies into the Maven repositories. It is also checks if the import statements in the ontology including transitive imports can be resolved. Therefore, it maintains an updated list of reference URIs to the ontology resources loaded to the repository.
This list follows the OASIS XMLCatalog standard which also specifies a technique for the automated replacement of external references in XML documents. A XML parser validates if defined replacement rules in an XMLCatalog apply to the references in the validated XML document and in case they apply, it replaces these references with the references defined in the XMLCatalog. In OntoMaven we use this automated replacement technology to replace the URI references to imported \textbf{external ontologies} with references to the \textbf{internal ontology artifacts}, which are locally managed in an OntoMaven repository after they have been loaded by the plug-in. This replacement approach avoids the continuous import and use of external ontologies during an OntoMaven development project. As an example the following catalog entry defines that the original reference to the imported ontology \texttt{example.owl} can be replaced by the URI referencing the stored ontology artifact in the local repository.

\begin{scriptsize}
\begin{verbatim}
<?xml version="1.0" encoding="UTF-8"?>
<catalog xmlns="urn:oasis:names:tc:entity:xmlns:xml:catalog">
    <system systemId="www.example.com/example.owl" uri="src/resource/owl"/>
</catalog>
\end{verbatim}
\end{scriptsize}

After the first loading of an ontology as repository artifact, including all transitive imports which are resolved and stored as dependent ontology artifacts, by the OntoMvnImport plug-in, the plug-in always checks if there is an ontology artifact listed in the XMLCatalog. If there is an existing reference to an ontology artifact, it will use it instead of any externally referenced ontology. A special situation is, if the import statement cannot be resolved, e.g. because the ontology is no longer existent under the given reference. In this case the plug-in notifies the user.

The following listing shows how the plug-in can be used in a OntoMaven POM. In the \texttt{configuration} it defines the input ontology and sets the \texttt{local} parameter to true, indicating that the ontology should be loaded to the local repository and that the local version of the ontology should be used.

\begin{scriptsize}
\begin{verbatim}
<build> <plugins> <plugin>
    <groupId>de.csw.ontomaven</groupId>
    <artifactId>OntoMvnImport</artifactId>
    <version>1.0-SNAPSHOT</version>
    <configuration>
        <owlfile>src/resource/reputation.owl</owlfile>
        <local>true</local>
    </configuration>
    <executions>
        <execution>
            <goals>
                <goal>owlimport</goal>
            </goals>
        </execution>
    </executions>
</plugin> </plugins> </build>
\end{verbatim}
\end{scriptsize}

\subsection{OntoMvnSvn}
The \texttt{OntoMvnSvn} plug-in provides ontology versioning support for OntoMaven. As discussed in section \ref{versioning} standard (code) versioning tools such as Concurrent Version System (CVS) and Subversion cannot be directly used, because version differences are only computed syntactically, but not semantically, as it is required for versioning interpreted knowledge models such as ontologies.

We therefore implemented an extension to Subversion called SVont\footnote{\scriptsize{\url{http://www.corporate-semantic-web.de/svont.html}}} \cite{conf/gi/Luczak-RoschCPRT10} which can compute semantic differences\footnote{\scriptsize{for the decription logic $\mathcal{EL}$}} and which can version ontologies. SVont supports typical Subversion commands such as \texttt{checkout}, \texttt{status}, \texttt{diff}, \texttt{commit}, and \texttt{info}. In the following the implementation of the commands \texttt{status} and \texttt{diff} are described in more detail.

For the \emph{status} command the OntoMvnSvn plug-in first does a repository checkout of the ontology to a temporal folder and then compares the repository version with the currently developed working version. The possible status results are \emph{identical} and \emph{changed}. This status information is used by the plug-in to either report \emph{ontology changed} or \emph{ontology is up-to-date}.

The \emph{diff} command first does a checkout of the committed repository version. It then computes the semantic difference to the working version. Therefore, the OntoMaven implementation is using the semantic difference compuation implemented by Svont. The OntoMvnSvn plug-in goal \texttt{diff} additionally performs a dependency analysis. It lists all dependencies, e.g. dependencies by domain and range properties and subclass relations as the following example shows:

\begin{scriptsize}
\begin{verbatim}
-------------------------- DIFF INFORMATION --------------------
Ontology File : ...\...\...\...\...\camera.owl
================== ACTUAL CHANGES ==========================
Axioms were added to the repository, or deleted from the working
copy.
SubClassOf(<www.xfront.com/owl/ontologies/camera/#Money>
owl:Thing)
Declaration(Class(<www.xfront.com/owl/ontologies/camera/#Money>))
=======================================================================
--------- MORE INFO --------------------------------
The above changes of the OWL classes are dependent on the following
axiom.
currency <------ DataProperty (Domain)
cost <------ ObjectProperty (Range)
--------------------------------------------------------------------
\end{verbatim}
\end{scriptsize}

\subsection{OntoMvnReport}

The plug-in is implemented as Maven report plug-in \footnote{\scriptsize{\url{http://docs.codehaus.org/display/MAVENUSER/Write+your+own+report+plugin}}}. The goal \texttt{site} of this plug-in creates four different documentations about the ontology - a general \emph{project documentation}, an \emph{ontology report summary}, a \emph{technical report}, and an \emph{ontology visualization}.

\begin{figure}
  \centering
  \includegraphics[width=7cm]{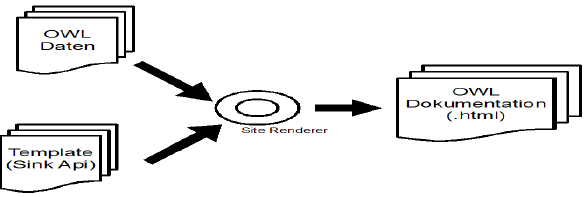}\\
  \caption{Simplified Diagram for the Report Rendering in the OntoMvnReport Plug-In }\label{fig:ReportRenderingDiagram}
\end{figure}

For rendering the ontology information into an HTML documentation it uses the Sink API \footnote{\scriptsize{\url{http://maven.apache.org/doxia/doxia/doxia-sink-api/}}} as illustrated in Figure \ref{fig:ReportRenderingDiagram}.

With the Maven Site Descriptor\footnote{\scriptsize{\url{http://maven.apache.org/plugins/maven-site-plugin/examples/sitedescriptor.html}}} the layout and content menus in OntoMaven reports can be adapted.

\begin{figure}
  \centering
  \includegraphics[width=2cm]{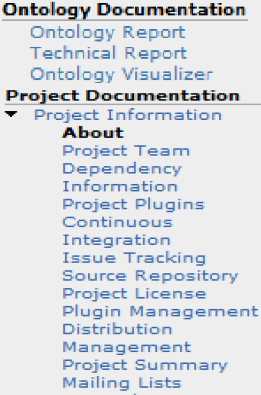}\\
  \caption{OntoMaven Project and Ontology Documentation Menu}\label{fig:ProjectDocumentation}
\end{figure}

The general \emph{ontology project documentation} is created from the description of the ontology artifact in the POM file, which includes project metadata about, e.g. the project, project team, dependencies, the plug-ins, issues, source repositories, licenses, ontology developers and their roles, etc. The following listing gives an example of typical project's metadata in a POM which is used in the project documentation which can be selected from the report menu (see figure \ref{fig:ProjectDocumentation}):

\begin{scriptsize}
\begin{verbatim}
<description>here's the descripton of an ontology </description>
<organization>
    <name>Corporate Semantic Web, Freie Universität Berlin</name>
    <url>www.corporate-semantic-web.de</url>
</organization>
<inceptionYear>2013</inceptionYear>
<licenses>
    <license>
        <name>LGPL-3.0</name>
        <url>www.gnu.org/licenses/lgpl.txt</url>
    </license>
</licenses>
<developers>
    <developer>
        <name>Adrian Paschke</name>
        <email>paschke@inf.fu-berlin.d</email>
        <organization>Corporate Semantic Web</organization>
        <organizationUrl>www.corporate-semantic-web.de/
        </organizationUrl>
        <roles>
            <role>developer</role>
        </roles>
    </developer>
</developers>
\end{verbatim}
\end{scriptsize}

\begin{figure}
  \centering
  \includegraphics[width=8cm]{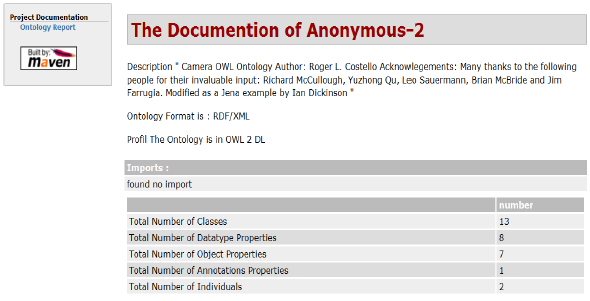}\\
  \caption{OntoMaven Ontology Report Summary}\label{fig:OntologySummary}
\end{figure}

The \emph{ontology report summary} is created by the goal \texttt{ontologyreport}. Figure \ref{fig:OntologySummary} shows an example ontology summarization which gives an overview about the general description, the format, the semantic profile, imported ontologies and a summary about the ontology's statistics (number of classes, datatype properties, object properties, etc.).

For the documentation of the ontology, the plug-in uses existing automated ontology documentation tools. We have integrated the SpecGen ontology documentation tool which creates a HTML page containing detailed information about the classes and the properties. We further extended SpecGen with various algorithms for creating structure based concept groupings. \cite{Coskun2012,DBLP:conf/womo/CoskunRTP11} These groupings are used as basis for a visual documentation of the ontology. To support this process of creating such concept groups for the documentation of ontologies we extended the SpecGen tool with an automatic concept grouping functionality\footnote{\scriptsize{\url{http://www.corporate-semantic-web.de/concept-grouping.html}}} and embedded it for the OntoMaven documentation.

\begin{figure}
  \centering
  \includegraphics[width=7cm]{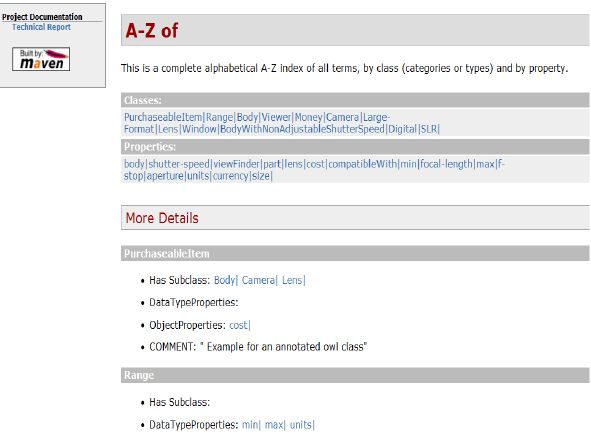}\\
  \caption{OntoMaven Technical Report}\label{fig:TechnicalReport}
\end{figure}

A more detailed insight is given by the \emph{technical ontology report}, which is created by the goal \texttt{technicalreport}. This goal produces a listing of classes and properties as shown in figure \ref{fig:TechnicalReport}. By clicking on a particular class or property the technical details about it are shown.

\begin{figure}
  \centering
  \includegraphics[width=8cm]{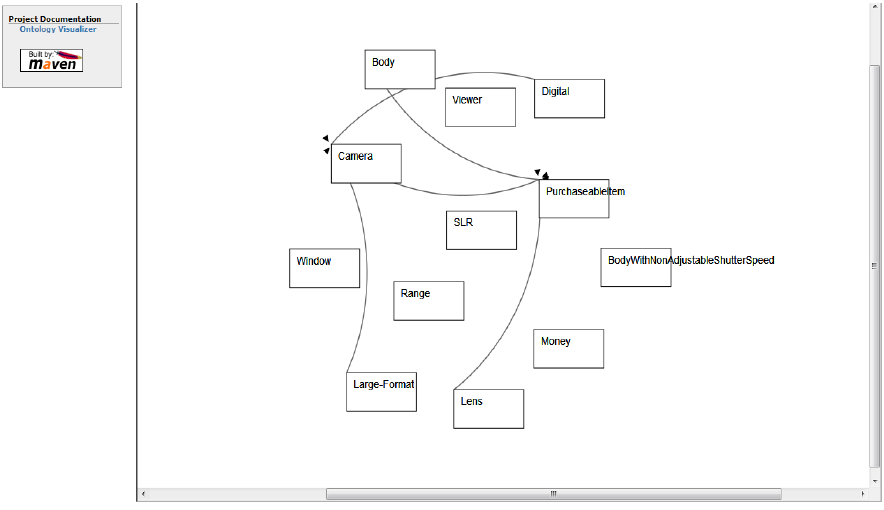}\\
  \caption{OntoMaven Ontology Visual Report}\label{fig:VisualReport}
\end{figure}

The goal \texttt{visualizer} produces a network visualization of the ontology concepts (classes) and its relations using different graph visualizations \footnote{\scriptsize{\url{http://www.corporate-semantic-web.de/ontology-modularization-framework.html}}}. Figure \ref{fig:VisualReport} gives a visualization example.

The following listing shows how to use the \texttt{OntoMvnReport} plugin in a project POM.xml

\begin{scriptsize}
\begin{verbatim}
<reporting><plugins><plugin>
<groupId>de.nbi.MvnOnt</groupId>
<artifactId>MvnOwlReport</artifactId>
<version>1.0-SNAPSHOT</version>
<reportSets>
    <reportSet>
        <configuration></configuration>
        <reports>
            <report>ontologyreport</report>
            <report>technicalreport</report>
            <report>visualizer</report>
        </reports>
    </reportSet>
</reportSets></plugin></plugins></reporting>
\end{verbatim}
\end{scriptsize}

The produced reports can be found in the Maven folder \texttt{target/site}.

\subsection{OntoMvnTest}

The \emph{OntoMvnTest} plug-in implements functionalities for the test phase. The plug-in executes the configured test using the goal \texttt{test}. It is also used internally in other phases such as the \texttt{package} goal. The plug-in implementation uses the Pellet reasoner\footnote{\scriptsize{\url{http://clarkparsia.com/pellet/}}} to execute the ontology test cases.

As default test suites the plug-in supports the W3C OWL Test Cases\footnote{\scriptsize{\url{http://www.w3.org/TR/owl-test/}}}. This test collection contain different types of test cases, such as a test that determines and returns the OWL sublanguage, tests for inconsistency checks, and entailment tests, which test if the intended conclusions (represented by an output ontology) are entailed in the input ontology model. For instance, the intended entailment test result values are \texttt{Entailment} (positive test result) or \texttt{NoEntailment} (negative test result).
The following listing shows how the plug-in can be used in a POM.xml.

\begin{scriptsize}
\begin{verbatim}
<build>
<plugins>
    <plugin>
        <groupId>de.csw.MvnOnt</groupId>
        <artifactId>MvnOwlTest</artifactId>
        <version>1.0-SNAPSHOT</version>
        <configuration>
            <owlfile>owl/1a.owl</owlfile>
        </configuration>
        <executions>
            <execution>
                <goals>
                    <goal>owltest</goal>
                </goals>
            </execution>
        </executions>
    </plugin>
    <plugin>
        <groupId>de.csw.MvnOnt</groupId>
        <artifactId>MvnOwlEntailment</artifactId>
        <version>1.0-SNAPSHOT</version>
        <configuration>
            <premise_file>owl/1a.owl</premise_file>
            <conclusion_file>owl/1aconclusion.
            owl</conclusion_file>
        </configuration>
        <executions>
            <execution>
                <goals>
                    <goal>owlentailment</goal>
                </goals>
        </execution>
    </executions>
</plugin></plugins></build>
\end{verbatim}
\end{scriptsize}

\section{Evaluation}
\label{evaluation}

OntoMaven is not a full ontology development tool as e.g. Protege, Swoop and Top Braid Composer, which provide a development user interface.
Instead OntoMaven's has its strength in the management of distributed ontology modules including support for reuse (transitive imports), dependency management and collaboration (semantic versioning).

Table \ref{ToolComparison} compares OntoMaven to the most used ontology development tools by its functional support in typical ontology engineering life cycles.

\begin{table}
\caption{Functional Comparison of OntoMaven with Ontology Development Tools} \label{ToolComparison}
\centering
\begin{scriptsize}
\begin{tabular}{|p{3cm}|p{2cm}|p{2cm}|p{2cm}|p{2cm}|}
  \hline
   & OntoMaven & Protege & Swoop & Top Braid Composer \\
  \hline
  Repositories & yes (local and remote) & yes (local and remote) & no & yes (by Allegro Graph 4 PlugIn) \\
  \hline
  Reuse (Import) & yes (dependency management) & yes & yes & yes \\
  \hline
  Collaboration Support (Versioning) & yes (semantic diff) & no & no & no \\
  \hline
  Documentation & yes (text and visual) & yes (text and visual) & yes (only text) & yes (text and visual in Maestro version) \\
  \hline
  Testing & yes & yes & yes & yes \\
  \hline
  Extensibility & yes & yes (many existing plugins) & yes & yes (commercial) \\
  \hline
\end{tabular}
\end{scriptsize}
\end{table}

\section{Conclusion}
\label{conclusion}

Apache Maven is a widespread and highly successful tool in Software Engineering for build automation and development project life cycle management. This paper has adapted the Maven approach and concepts for Knowledge Engineering in (agile) ontology development. The contribution is a new design artifact called OntoMaven which has been implemented as a proof-of-concept implementation.

OntoMaven uses a Project Object Model (POM) XML file to describe the ontology project being developed, its dependencies on other external ontology modules, the development life cycle order, directories, and required plug-ins. It comes with pre-defined targets for performing certain well-defined tasks in typical agile ontology development phases.

OntoMaven dynamically downloads distributed ontologies and Maven plug-ins from one or more OntoMaven remote repositories and stores them in a local cache, the local repository. This local cache of downloaded artifacts can be further updated with ontology artifacts created by local projects as well as developed artifacts can be uploaded to remote OntoMaven repositories, so that they can be shared and reused.

OntoMaven is built using Maven's plugin-based architecture that allows it to make use of any application controllable through standard input. The proof-of-concept of OntoMaven implements several useful plugins which interface with existing ontology development tools and functionalities such as the plug-ins \emph{OntoMvnImport}, \emph{OntoMvnSVN}, \emph{OntoMvnReport}, and \emph{OntoMvnTest}.

In future research we plan to study the benefits of the OntoMaven support for ontology engineers in real agile ontology development projects and compare it with other development approaches. We plan to quantify the results on the basis of ontology development costs (in person months) and compare them against the estimated cost models (based on our work on Agile ONTOCOM 2\footnote{\scriptsize{\url{http://www.corporate-semantic-web.de/corporate-ontology-engineering.html}}} \cite{PA13}).

In our aspect-oriented ontology development research \cite{SchPa13} we plan to make use of OntoMaven for the distributed management of ontology modules and their description as \emph{aspect-oriented ontology artifacts} in the POM.

\section{Acknowledgements}
This work has been partially supported by the InnoProfile project "Corporate Semantic Web" funded by the German Federal Ministry of Education and Research (BMBF).

\bibliographystyle{plain}
\bibliography{ontomaven}

\begin{thebibliography}{10}

\bibitem{DBLP:conf/wetice/Auer06}
S{\"o}ren Auer.
\newblock The rapidowl methodology--towards agile knowledge engineering.
\newblock In {\em WETICE}, pages 352--357, 2006.

\bibitem{CLHP2009a}
G.~Coskun, M.~Luczak-R\"osch, R.~Heese, and A.~Paschke.
\newblock Applying ontology modularization for corporate ontology engineering.
\newblock In {\em Proceedings of the Intl. Conference on Semantic Systems
  (I-SEMANTICS 2009)}, pages 669--674, Graz, Austria, Sep 2009.

\bibitem{Coskun2012}
G{\"o}khan Coskun, Mario Rothe, and Adrian Paschke.
\newblock Ontology content ``at a glance''.
\newblock In M~Donnelly and G~Guizzardi, editors, {\em Proceedings of the 7th
  International Conference on Formal Ontology in Information Systems}, pages
  147--159, Graz, Austria, 2012. IOS Press.

\bibitem{DBLP:conf/womo/CoskunRTP11}
G{\"o}khan Coskun, Mario Rothe, Kia Teymourian, and Adrian Paschke.
\newblock Applying community detection algorithms on ontologies for identifying
  concept groups.
\newblock In {\em WoMO}, pages 12--24, 2011.

\bibitem{DeNicola:2009:SEA:1467085.1467122}
Antonio De~Nicola, Michele Missikoff, and Roberto Navigli.
\newblock A software engineering approach to ontology building.
\newblock {\em Inf. Syst.}, 34(2):258--275, April 2009.

\bibitem{KhiMue10}
M.~Rahamatullah Khondoker and Paul Mueller.
\newblock Comparing ontology development tools based on an online survey.
\newblock In {\em Proceedings of the World Congress on Engineering 2010 (WCE
  2010)}, 2010.

\bibitem{kilic13}
Onur Kilic.
\newblock Erweiterung von maven zur toolbasierten verwaltung von
  ontologiemodulen, 2013.

\bibitem{DBLP:journals/corr/abs-1204-5093}
Christoph Lange, Oliver Kutz, Till Mossakowski, and Michael Gr{\"u}ninger.
\newblock The distributed ontology language (dol): Ontology integration and
  interoperability applied to mathematical formalization.
\newblock {\em CoRR}, abs/1204.5093, 2012.

\bibitem{conf/gi/Luczak-RoschCPRT10}
Markus Luczak-Rösch, Gökhan Coskun, Adrian Paschke, Mario Rothe, and Robert
  Tolksdorf.
\newblock Svont - version control of owl ontologies on the concept level.
\newblock In Klaus-Peter Fähnrich and Bogdan Franczyk, editors, {\em GI
  Jahrestagung (2)}, volume 176 of {\em LNI}, pages 79--84. GI, 2010.

\bibitem{RoeHe09}
Markus Luczak-Rösch and Ralf Heese.
\newblock Managing ontology lifecycles in corporate settings.
\newblock In Tassilo Pellegrini, Sóren Auer, Klaus Tochtermann, and Sebastian
  Schaffert, editors, {\em Networked Knowledge - Networked Media}, volume 221
  of {\em Studies in Computational Intelligence}, pages 235--248. Springer
  Berlin Heidelberg, 2009.

\bibitem{DBLP:journals/ijait/PaschkeB11}
Adrian Paschke and Harold Boley.
\newblock Rule responder: Rule-based agents for the semantic-pragmatic web.
\newblock {\em International Journal on Artificial Intelligence Tools},
  20(6):1043--1081, 2011.

\bibitem{DBLP:conf/icpw/PaschkeBKC07}
Adrian Paschke, Harold Boley, Alexander Kozlenkov, and Benjamin~Larry Craig.
\newblock Rule responder: Ruleml-based agents for distributed collaboration on
  the pragmatic web.
\newblock In {\em ICPW}, pages 17--28, 2007.

\bibitem{PaCoHe10}
Adrian Paschke, Gökhan Coskun, Ralf Heese, Markus Luczak-Rösch, Radoslaw
  Oldakowski, Ralph Schäfermeier, and Olga Streibel.
\newblock Corporate semantic web: Towards the deployment of semantic
  technologies in enterprises.
\newblock In Weichang Du and Faezeh Ensan, editors, {\em Canadian Semantic
  Web}, pages 105--131. Springer US, 2010.

\bibitem{PA155}
Adrian Paschke, G{\"o}khan Coskun, Dennis Hartrampf, Ralf Heese, Markus
  Luczak-R{\"o}sch, Mario Rothe, Radoslaw Oldakowski, Ralph Sch{\"a}fermeier,
  and Olga Streibel.
\newblock Realizing the corporate semantic web: Prototypical implementations.
\newblock TR-B-10-05:1--49, 02/2010 2010.

\bibitem{PA13}
Adrian Paschke, G{\"o}khan Coskun, Harasic Marko, Ralf Heese, , Radoslaw
  Oldakowski, Ralph Sch{\"a}fermeier, Olga Streibel, Kia Teymourian, and
  Alexandru Todor.
\newblock Corporate semantic web report vi: Validation and evaluation.
\newblock TR-B-13-01:1--64, 01/2013 2013.

\bibitem{SchPa13}
Ralph Sch{\"a}fermeier and Adrian Paschke.
\newblock Towards a unified approach to modular ontology development using the
  aspect-oriented paradigm.
\newblock In {\em 7th International Workshop on Modular Ontologies (WoMO
  2013)}, 2013.

\bibitem{DBLP:conf/ruleml/WeigandP12}
Hans Weigand and Adrian Paschke.
\newblock The pragmatic web: Putting rules in context.
\newblock In {\em RuleML}, pages 182--192, 2012.

\end{thebibliography}

\end{document}